\documentclass{JHEP3} 

\usepackage{epsfig}

\title{Perturbative and non-perturbative aspects of the\\
two-dimensional string/Yang-Mills  correspondence}
\author{Simone Lelli, Michele Maggiore and Anna Rissone\\
D\'epartement de Physique Th\'eorique, Universit\'e de Gen\`eve,\\
24 quai Ernest-Ansermet, CH-1211 Gen\`eve 4\\
E-mail: \email{simone.lelli, michele.maggiore,
anna.rissone@physics.unige.ch}}
\abstract{It is known that  YM$_2$ with gauge group $SU(N)$ 
  is equivalent to a  string theory with coupling $g_s=1/N$, 
  order by order  in the $1/N$ expansion.
  We show how this results can be obtained
  from the bosonization of the fermionic formulation of YM$_2$,
  improving on results in the literature,
  and we examine a number of non-perturbative
  aspects of this string/YM correspondence. 
  We find contributions to the YM$_2$ partition function 
  of order $\exp\{ -kA/(\pi\a 'g_s)\}$ with $k$ an
  integer and $A$ the area of the target space, which would 
  correspond, 
  in the string interpretation, to D1-branes. Effects which could be
  interpreted as D0-branes are instead stricly absent, suggesting a
  non-perturbative structure typical of type~0B string theories.
  We discuss effects from the YM side that are interpreted in terms
  of the  stringy  exclusion principle of Maldacena and Strominger.
  We also find numerically an interesting phase structure, with
  a region where YM$_2$ is described by a perturbative string theory 
  separated from
  a region where it is described by a topological string theory.
}
\preprint{UGVA-DPT 2002/10-1103}

\renewcommand\({\left(}
\renewcommand\){\right)}
\renewcommand\[{\left[}
\renewcommand\]{\right]}

\newcommand{\ra}{\rightarrow}

\def\lsim{\raise 0.4ex\hbox{$<$}\kern -0.8em\lower 0.62
ex\hbox{$\sim$}}

\def\gsim{\raise 0.4ex\hbox{$>$}\kern -0.7em\lower 0.62
ex\hbox{$\sim$}}

\newcommand\eq[1]{eq.~(\ref{#1})}
\newcommand\eqs[2]{eqs.~(\ref{#1}) and (\ref{#2})}

\newcommand\pa{\partial}

\newcommand\ee{\end{equation}}
\newcommand\be{\begin{equation}}
\def\bea{\begin{eqnarray}}
\def\eea{\end{eqnarray}}
\newcommand\ees{\end{eqnarray}}
\newcommand\bees{\begin{eqnarray}}

\newcommand{\bnd}{B_n^{\dagger}}
\newcommand{\bmd}{B_m^{\dagger}}
\newcommand{\fvac}{|0\rangle_F}
\newcommand{\vac}{|0\rangle}
\newcommand{\bt}{\tilde{b}}
\newcommand{\ct}{\tilde{c}}
\newcommand{\at}{\tilde{\a}}

\def\dslash{\not{\hbox{\kern-2pt $\partial$}}}
\def\Dslash{\not{\hbox{\kern-4pt $D$}}}
\def\pslash{\not{\hbox{\kern-2.3pt $p$}}}

\def\D{\Delta}

\def\a{\alpha}
\def\b{\beta}

\def\th{\theta}

\def\g{\gamma}

\def\l{\lambda}

\def\d{\delta}

\newcommand\fverb{\setbox\pippobox=\hbox\bgroup\verb}
\newcommand\fverbdo{\egroup\medskip\noindent%
			\fbox{\unhbox\pippobox}\ }
\newcommand\fverbit{\egroup\item[\fbox{\unhbox\pippobox}]}
\newbox\pippobox

\begin{document}

\section{Introduction}

In the early nineties Gross \cite{Gross} and Gross and Taylor \cite{GT}
showed that two-dimensional pure YM theory with gauge group $SU(N)$
on a euclidean manifold of arbitrary topology is equivalent,
order by order in the large $N$ expansion, to a string theory with 
coupling $g_s=1/N$ (see e.g. refs.~\cite{CMR}--\cite{BDP} for further
developements).

In the light of the recent advances on  string/YM
correspondence it is interesting to go back
to this result, for a number of reasons. First, in this
two-dimensional setting the correspondence can be  {\em
  proven} mathematically, at least at the level 
of perturbation theory. This comes from the
remarkable fact that the partition function of YM$_2$ on an arbitrary euclidean
manifold, with  gauge group $U(N)$ or $SU(N)$ and $N$ generic, can be computed
exactly. Second, this theory has no space-time
supersymmetry, suggesting that supersymmetry is not a necessary
ingredient for a string/YM correspondence to 
hold-- a fact of obvious importance if
one hopes to apply the correspondence to QCD.
And finally, in the years
after refs.~\cite{Gross,GT} came out, D-branes have been introduced
and the understanding of
non-perturbative string theory has developed greatly, so it becomes
possible to ask whether this correspondence holds even beyond
perturbation theory. 

In this paper we consider some aspects of the relation between
YM$_2$ and string theory. In sect.~\ref{sect2} we  briefly recall the
main results of refs.~\cite{Gross,GT,CMR,CMR2}, 
where it is shown that the $1/N$
expansion of YM$_2$ can be interpreted geometrically in terms of a
theory of maps from a two-dimensional world-sheet to a two-dimensional
target space. We also  recall the result of Minahan 
and Polychronakos~\cite{Min}, who showed
that this expansion can be elegantly summarized in terms 
of a ``string field theory''
Hamiltonian, i.e. a Hamiltonian containing operators that create and
destroy strings with a given winding over the cycles of the
target manifold.
This Hamiltonian, for $U(N)$, consists of a term  $O(1)$ plus a term $O(1/N)$
(for $SU(N)$, there is also a term $O(1/N^2)$) and
all other perturbative corrections to it in powers of
$1/N$ are exactly zero; the full and complicated  $1/N$
expansion of the YM$_2$
partition function is completely reproduced by the
expansion of the exponential of this Hamiltonian, traced over a
multistring Fock space. Thus this Hamiltonian summarizes very compactly all the
perturbative expansion, and is useful to clarify the physical
meaning of this
two-dimensional string-YM correspondence. 

In sect.~\ref{sect3} we show how this Hamiltonian
can be rigorously derived from a bosonization of the
fermionic formulation of YM$_2$. The idea behind the computation has
been described by Douglas~\cite{Dou1,Dou2}. However, strictly speaking
the derivation of refs.~\cite{Dou1,Dou2} only shows that the
Hamiltonian of Minahan and Polychronakos
is obtained as the leading term in the large $N$ limit, while we
will see explicitly that it is exact, i.e. 
all its further perturbative corrections in $1/N$ vanish. This
completes a simple and rather elegant proof of the perturbative
correspondence.

In sect.~\ref{sect4} we examine some non-perturbative aspects of the
correspondence. The expansion of the YM$_2$ partition function at large
$N$ has in fact also terms $e^{-O(N)}$, already noted by Gross~\cite{Gross},
which should match with contribution $e^{-O(1/g_s)}$ of the
corresponding string theory, if the correspondence holds even beyond
the perturbative level.
Indeed we will find that, from the YM side,
there is a set of contributions  proportional to
$e^{-kA/(\pi\a ' g_s)}$, with $k$ an integer,  $A$ the target space area,
$\a'$ the string tension of the string theory (fixed by the 't~Hooft
coupling of the YM theory, see below) and $g_s=1/N$.
The factor $1/g_s$ at the exponent 
is suggestive of D-branes. More precisely,
the proportionality to the
area of the target space is just what one would expect from
$D1$-branes in this string theory. In fact, the strings corresponding
to YM$_2$ have the peculiar properties of having no
foldings~\cite{Gross}, i.e. their world-sheet area is an integer times
the target space area. It is then natural to expect the same for
the $D1$-branes, and indeed the factor $kA$ in the exponent can be
interpreted as the world-sheet area of a $D1$-brane wrapping $k$ times
over the target space without foldings, and $\tau_1 =1/(\pi\a ' g_s)$ 
can be interpreted as the 
$D1$-brane tension.
We will see that instead there is no
effect that has an interpretation in terms of $D0$-branes. We therefore
find a non-perturbative structure typical of type~B 
string theories: $p$-branes
with $p$ even are absent and with $p$ odd are present.

We will also find that a  non-perturbative string effect as the
stringy exclusion principle of Maldacena and Strominger~\cite{MaldaStro}
appears from the YM$_2$ side, as a very simple consequence of the
fermionic formulation of YM$_2$. We will then discuss our attempts
to evaluate numerically the non-perturbative effects in YM$_2$, and we
will find an interesting structure in the plane $(g_s,a)$, where $a=\l A/2$,
$\l =e^2N$   is the 't~Hooft coupling of the YM theory
and $A$ is the area of the target space.

\section{The large-$N$ expansion of YM$_2$}\label{sect2}

We consider pure Yang-Mills theory on a two dimensional euclidean
manifold ${\cal M}$ of arbitrary topology, with gauge group $U(N)$ or
$SU(N)$ and charge $e$.  The partition function can be written
as a sum over all representations $R$ of the gauge
group~\cite{Migdal,Rusakov} 
\be\label{ZYM}
Z_{\rm YM}\equiv \int [{\cal D}A^{\mu}]\, \exp
\{ -\frac{1}{4e^2}\int_{\cal M}d^2x\sqrt{g}\,
{\rm Tr} F^{\mu\nu}F_{\mu\nu} \}=
\sum_R\, ({\rm dim }R)^{2-2G}e^{-\frac{\l A}{2N}C_2(R)}\, ,
\ee
where $G$ is the genus of ${\cal M}$, $A$ its area, $\l =e^2N$ is the
't~Hooft coupling, to be held fixed in the large $N$ expansion, and
$C_2(R)$ is the quadratic Casimir in the representation $R$.

The representations $R$ of $U(N)$ or $SU(N)$ are given by the Young
diagrams with $m$ rows, with $m\leq N$ for $U(N)$ and $m<N$ for
$SU(N)$. Denoting by $h_i$, $i=1,\ldots m$, the number of boxes in the 
$i$-th row (with $h_N=0$ for $SU(N)$) and by $c_j$ the number of boxes in the 
$j$-th column, the quadratic Casimir can be written as~\cite{Gross} 
\bees\label{C2}
C_2^{U(N)}(R)&=&N n +\tilde{C}(R)\, ,\\
C_2^{SU(N)}(R)&=&N n +\tilde{C}(R)-\frac{n^2}{N}\, ,\label{C2SUN}
\ees
where $n=\sum_{i=1}^N h_i=\sum_{j=1}^{\infty} c_j$ 
is the total number of boxes in the Young diagram,
and
\be\label{Ctilde}
\tilde{C}(R)=\sum_{i=1}^Nh_i^2-\sum_{j=1}^{\infty}c_j^2\, .
\ee
Observe that each of the $h_i$ takes values in the range
$0\leq h_i <\infty$ and its index $i$ takes the values $i=1,\ldots , N$,
i.e. the number of rows is limited by $N$ (with $h_N=0$ for $SU(N)$)
but the rows can be arbitrarily long. Instead $0\leq c_j\leq N$, with
$j=1,\ldots ,\infty$, corresponding to the fact that the length of the
columns is limited by $N$ (by $N-1$ for $SU(N)$) but the number of columns
is arbitrary. This asymmetry  between the $h_i$ and the $c_j$ is
important when one considers non-perturbative effects, as we shall see.

The dimension of the representation, dim~$R$, has also a closed form in
terms of the $h_i$~\cite{Gross} and therefore one has a very explicit
expression for the partition function, which can be expanded in powers
of $1/N$.

The beautiful result of Gross~\cite{Gross} 
is that, order by order in  $1/N$,
all terms in the expansion of
the logarithm of $Z_{\rm YM}$ can be interpreted geometrically as a sum
of contributions due to 
maps from a two dimensional world-sheets to the target space
${\cal M}$, or, more precisely, as a sum over all possible
branched coverings of
${\cal M}$, so that one can identify $\log Z_{\rm YM}$
 with the
partition function of a string theory with coupling $g_s=1/N$ and
string tension $\a' \sim 1/\l$ (recall that in two dimensions the electric
charge $e$ has dimensions of mass, so $\l =e^2N$ is a mass squared):
\be\label{log}
\log Z_{\rm YM}[G,A,\l ,N]=Z_{\rm
  string}\[ g_s=\frac{1}{N},\a'=\frac{1 }{\pi\l }\]\, .
\ee
The details of this identification,
fully worked out in refs.~\cite{GT,Gross2,CMR}, are quite
intricated, but basically
one finds that the terms in the expansion of the left-hand side
are weighted by a factor
$\exp (-n\l A/2 )$, with $n$ a summation index which is
interpreted as the number of sheets of the covering, so that the
factor $nA$ has the geometric interpretation of the area of the
world-sheet of a string which has no foldings, 
and $\l /2$ is then identified with the string tension
$1/(2\pi\a ')$; the
identification of $g_s$ with $1/N$ comes from the presence of factors
$N^{\chi}$, with $\chi$ equal to the Euler characteristic of the
branched covers
(which includes the contribution of the singularities of the branched  
cover); furthermore, the overall
coefficient associated to each contribution of the sum
(i.e. to each branched cover) turns out to be 
related to  the number of topologically inequivalent
maps from the given branched cover to the target space. Therefore 
$\log Z_{\rm YM}$ has a full geometric interpretation,
and has the structure of the partition function of a theory of maps.

The relation $Z_{\rm YM}=\exp ( Z_{\rm string})$ is of course the same
relation that one has between the partition function of a first
quantized particle, $Z_{S^{1}}=\int Dx^{\mu}\, e^{-S}$, computed
integrating over
all  trajectories $x^{\mu}(\tau )$ with the topology of the circle, 
and the partition function of the
corresponding field theory, $Z_{\rm vac}=\exp (Z_{S^{1}})$. So
\eq{log}  means 
that YM$_2$ is rather a string field theory. 

This point becomes
evident when one realizes that the whole complicated $1/N$ expansion can be
summarized very concisely in terms of a 
Hamiltonian acting on a Fock space generated by operators that create
and destroy strings~\cite{Min}. To understand this, one observes first of all
that  the  YM$_2$ partition function on a 
surface of arbitrary genus can be obtained from the partition function
on the cylinder by using the gluing property~\cite{Witym2}, so we can
limit ourselves to the partition function on a cylinder of
circumference $L$ and length $T$.
To quantize YM$_2$ on a cylinder  one chooses the gauge
$A_0=0$ and is therefore left with wave-functionals
$\Psi[A_1(x)]$. The constraint obtained varying with respect to $A_0$
imposes that $\Psi[A_1(x)]$ actually depends only on the holonomy
$U=P\exp[i \int_0^Ldx A_1]$. The Hilbert space of states can
therefore be labelled by the holonomies, $|U\rangle$~\cite{CMR}.

We then introduce the Fock space generated by the operators $\a_n$,
with $[\a_n,\a_m ]=n\d_{n+m}$. Physically $\a_n$ with $n>0$ destroys a
string winding $n$ times in the clockwise direction around the
cylinder and $\a_{-n}$ creates it. We also introduce a second set
$\tilde{\a}_n$ creating and destroying strings winding in the
counterclockwise direction. A generic multistring state is therefore
of the form~\cite{CMR}
\be\label{multi}
|\vec{k},\vec{l}\,\rangle = \prod_{i>0}(\a_{-i})^{k_i}
\prod_{j>0}(\tilde{\a}_{-j})^{l_j} |0\rangle\, .
\ee
Now we consider the YM$_2$ partition function on a cylinder, with holonomies 
$U_1,U_2$ at the boundaries,
\be\label{cylinder}
Z_{\rm cyl} = \langle U_1 | e^{-HT} | U_2\rangle\, ,
\ee
where $H$ is the YM$_2$ Hamiltonian. On the one hand, this can be
computed exactly in closed form, similarly to (\ref{ZYM}). On the
other hand, we can rewrite it as
\be\label{setcompl}
Z_{\rm cyl} = \sum_{s,s'} 
\langle U_1 |s\rangle \langle s| e^{-HT} |s'\rangle
\langle s'|U_2\rangle\, ,
\ee
where $|s\rangle ,|s'\rangle$ are a complete set of multistring states of
the type (\ref{multi}). The matrix elements $\langle U |s\rangle $
are fixed requiring that \eq{setcompl} reproduces the dependence of
$Z_{\rm cyl}$ on the holonomies. When the state $|s\rangle$ is
constructed only from
operators $\a_{-n}$ (or  only from $\tilde{\a}_{-n}$) the result is
especially simple\cite{CMR}\footnote{For the most general case, see
  ref.~\cite{CMR}, sect.~4.7.1.},
\be
\langle U |\vec{k}\rangle  =
\prod_{j=1}^{\infty} ({\rm Tr }\,  U^j)^{k_j}\, .
\ee
Then the Hamiltonian $H$ in the string
basis is fixed  requiring that, when
inserted into \eq{setcompl}, it reproduces 
the full $1/N$ expansion of the cylinder amplitude,
and for $SU(N)$ it turns out to be~\cite{Min}
\begin{eqnarray}
H & = & \frac{\lambda L}{2}\left\{(\mathcal{N} + \tilde{\mathcal{N}})
- \frac{1}{N^2} (\mathcal{N} 
- \tilde{\mathcal{N}})^2 + \frac{1}{N} \sum_{n , l > 0}
\left(\a_{-n-l} \a_n \a_l \, + \, \a_{-n} \a_{-l}
\a_{n+l}\right) + \right.\nonumber\\ 
& + & \left.\frac{1}{N} \sum _{n , l > 0}\left( 
 \tilde{\a}_{-n-l}\tilde{\a}_n
\tilde{\a}_l \, + \, \tilde{\a}_{-n} \tilde{\a}_{-l}
\tilde{\a}_{n+l} \right)\right\}\, ,\label{ham-min}  \label{H}
\end{eqnarray}
where $\mathcal{N} = \sum_{n = 1}^{\infty} \a_{-n} \a_n$.

Eq.(\ref{H}) shows in the clearest way that YM$_2$ is equivalent to a string
field theory, since all matrix elements can be computed in terms of
a Hamiltonian and a Fock space  constructed 
using operators that create and destroy
strings with a given winding number around the cylinder.

Considerable effort has gone into trying to reproduce $Z_{\rm
string}$ in \eq{log} from the path integral over a suitable string
action~\cite{CMR,CMR2,Horava}, in order to make contact with
the standard 
first-quantized formalism of string theory. It appears, however, that
if such a formulation exists at all, it is  very complicated, except in the
limit of vanishing target-space area, $A\ra 0$, where one finds a
topological string theory. On the other hand,
at least at the perturbative level, a first quantized formulation
is not really  necessary,
since in this case we are in the rather unique
situation of having already at our disposal a second quantized string
theory, defined by the Hamiltonian (\ref{H}), which furthermore has an
extremely simple form, with just a free piece plus cubic and quartic
interaction terms, and, at least at the level of perturbation theory,
contains all the informations that we need on the stringy
description of YM$_2$.

The Hamiltonian (\ref{H}) was first found~\cite{Min} as a sort of
bookkeeping device that summarizes the whole
$1/N$ expansion of $Z_{\rm YM}$. 
One can ask whether it can be derived directly from the YM$_{2}$
action, shortcutting the  highly elaborated procedure of the $1/N$
expansion. In fact this is possible, if one starts from the fermionic
formulation of YM$_2$ and then bosonizes it, as was understood by
Douglas~\cite{Dou1,Dou2}. Actually, while we can see,
following refs.~\cite{Dou1,Dou2}, that the
Hamiltonian $H$ emerges from this bosonization procedure
in the large $N$ limit,
a little more care is needed to make sure that $H$ in \eq{H} is
reproduced exactly, with no further subleading
term in $1/N$. Since
the great power of the Hamiltonian (\ref{H}) is just that it is exact at
all orders in $1/N$, we find useful in the next section to perform the
calculation carefully, 
verifying explicitly the cancellation of the subleading terms.
We will also find the  expression for the $U(N)$ Hamiltonian,
which is not correctly given in the literature.

\section{The string Hamiltonian}\label{sect3}

\subsection{Fermionic representation of YM$_2$}

The starting point is the description of YM$_2$
in terms of free non-relativistic fermions~\cite{Parisi,Dou1}
(see also ref.~\cite{CMR} for review). 
We have seen that in the functional
Schroedinger equation the wave-functional $\Psi$ depends only on the
holonomies $U$; by gauge invariance, it must indeed be a class
function, i.e. $\Psi[U]=\Psi[gUg^{-1}]$ with $g$ a gauge
transformation, $g\in U(N)$ or $SU(N)$. Class functions depends only
on their value on the maximal torus, whose elements can be
parametrized as $\rm{diag}(e^{i\th_1},\ldots ,e^{i\th_N})$ (with the
further constraint $\sum_i\th_i=0$ for $SU(N)$). Then 
$\Psi=\Psi [\vec{\th}\, ]$ and, by Weyl symmetry, is symmetric under
exchange of any two $\th_i$. The inner product on class function is
fixed by the invariant measure over the group and is
\be
(\Psi ,\Psi )=\int \prod d\th_i\, \tilde{\D} (\vec{\th}\, )^2 |\Psi
(\vec{\th}\, )|^2\, ,
\ee
with $\tilde{\D} =\prod_{i<j}\sin [(\th_i-\th_j)/2]$. 
The YM$_2$ Hamiltonian
acting on $\Psi[\vec{\th}]$ is, for $U(N)$,
\be\label{Hth}
H_{U(N)}=\frac{e^2L}{2}\frac{1}{\tilde{\D}[\vec{\th}\, ]}
\[ \sum_i \( -\frac{d^2}{d\th_i^2}\) -\frac{N}{12} (N^2-1)\]
\tilde{\D}[\vec{\th}\, ]\, ,
\ee
while $H_{SU(N)}=H_{U(N)}-(e^2L/2) Q^2/N$, with $Q$ the $U(1)$
generator, see below.
We can therefore work with a new wave-functional $\psi [\vec{\th}\, ]=
\tilde{\D}[\vec{\th}\, ]\Psi[\vec{\th}\, ]$, in terms of which both the
inner product and the functional Schroedinger equation are those of a
free theory. However, since $\Psi$ is symmetric and $\tilde{\D}$
antisymmetric, $\psi$ is antisymmetric, and the YM$_2$ theory 
with gauge group $U(N)$ or $SU(N)$ is
therefore reduced to the quantum mechanics of $N$ free non-relativistic
fermions, with
each fermion described by a coordinate
$\th_i$, $i=1,\ldots N$ and therefore living
on the circle, and with the further constraint $\sum_i\th_i=0$ for
$SU(N)$.
\EPSFIGURE[h]{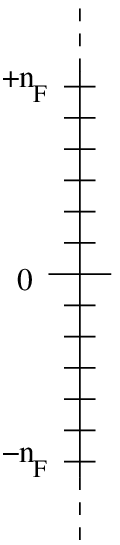}{\label{fig1} The filled fermionic levels in the
  ground state of $SU(N)$ YM$_2$ (when $N$ is odd).}
The generic state of this fermionic system is labelled as
\be
|n_1,\ldots n_N\rangle
\ee
with $n_i\in {\cal Z}$ and $n_1>n_2>\ldots n_N$, by the exclusion
principle. The energy of such a state is read from \eq{Hth} and is
\be\label{E}
E_{U(N)}=\frac{e^2L}{2}\[\sum_{i=1}^N n_i^2 -\frac{N}{12} (N^2-1) \]
\ee
while the $U(1)$ charge is easily seen to be
$Q=\sum_{i=1}^N n_i$.
The ground state, restricting for simplicity to
$N$ odd,\footnote{The analysis that we will discuss can be repeated
  with very minor modifications for $N$ even. In order not to burden
  all arguments, repeating them for $N$ even and $N$ odd, we will just
  restrict to $N$ odd. No interesting new feature appears for $N$ even.} 
is obtained filling all levels from
$-n_F$ to $n_F$, see fig.~\ref{fig1}, with the Fermi surface at 
\be
n_F=\frac{N-1}{2}\, .
\ee

For this state
\be\label{E0}
\sum_{i=1}^N n_i^2 = 2\sum_{i=1}^{n_F} i^2 = \frac{N}{12} (N^2-1)
\ee
and therefore the energy (\ref{E}) is zero. 
Each fermionic configuration $\{ n_i\}$ corresponds to a Young diagram
with rows of length~\cite{CMR}
\be\label{Young}
h_i=n_i+i-1-n_F\, ,
\ee
and therefore the partition
function (\ref{ZYM}) is immediately rewritten as a sum over all
fermionic configurations. For $U(N)$, the representation is labelled
also by the $U(1)$ charge.
For $SU(N)$, 
two fermionic configurations correspond to the same
Young diagrams if they are related by a global shift of the $n_i$,
$n_i\ra n_i+b$, $b\in {\cal Z}$. We can use this freedom to set
$n_N=-n_F$. 

Even if the total number of fermions, $N$, is fixed for a given $U(N)$
or $SU(N)$ YM theory,
it turns out to be  convenient to
introduce a second quantization formalism, defining $B_n$ 
(with $n\in {\cal Z}$) as the
operator that destroys a fermion in the state $|n\rangle$ and 
$B_n^{\dagger}$ as the creation operator, with $\{ B_n,\bmd
\}=\d_{n,m}$. The number operator is therefore 
$\hat{N}=\sum_{n=-\infty}^{\infty} \bnd B_n$. 
The vacuum $\vac$ is defined by $B_n\vac=0$ for all $n$. 
However, it is not a state of the $U(N)$ or $SU(N)$ theory, 
since it does not have $N$ occupied levels.
We instead define the Fermi vacuum $\fvac$ from
\bees
B_n\fvac =0 &\vspace{2cm} & {\rm if }\,\, |n|>n_F\\
\bnd\fvac =0 &\vspace{2cm} & {\rm if }\,\, |n|\leq n_F\, 
\ees
In other words, 
\be
\fvac =\prod_{n=-n_F}^{n_F} \bnd\vac\, .
\ee
We use $B_n,\bnd$ to define operators in which the mode number is
measured with reference to the two Fermi surfaces at $n=\pm n_F$:
\be\label{bc}
\left. \begin{array}{ll}
       c_n = B_{n_F+1-n}^{\dagger} \\
       b_n = B_{n_F+1+n}
       \end{array}
\hspace{1cm} \right\}  |n| \leq n_F
\ee
and 
\be\label{bct} 
\left. \begin{array}{ll}
       \tilde{c}_n = B_{-(n_F+1)+n}^{\dagger} \\
       \tilde{b}_n = B_{-(n_F+1)-n}
       \end{array}
\hspace{1cm} \right\}  |n| \leq n_F\, .
\ee
If we would  extend the definitions of $b_n,c_n$ and
$\tilde{b}_n,\tilde{c}_n$ at $|n|>n_F$ then
they would not be independent, since e.g. the same operator $B_m$ would be
assigned both to one of the $b_n$ and to one of the $\tilde{b}_n$;
we find simpler to put a
cutoff on the mode number $n$ and work with independent 
quantities.\footnote{In principle, one might 
decide to use an asymmetric cutoff; for instance, in $b_n,
\tilde{b}_n$ all we
really need is a lower bound on $n$ for both $b_n$ and
$\tilde{b}_n$, so that the $b_n$ and the $\tilde{b}_n$
do not 'collide' with each other;
an upper bound like $n\leq n_F$ in $b_n,\tilde{b}_n$ or a lower bound
for $n$ in $c_n,\tilde{c}_n$ are not
necessary. It is however sligthly simpler to put the cutoff
symmetrically, which means that we forbid very high
excitations like $\bnd\vac$ with
$n> N$.
As we will see, at all orders in perturbation theory in
$1/N$, these definitions are equivalent.}
In terms of the $B_n,\bnd$ the cutoff is such that are included all
operators $B_n,\bnd$ with $-N \leq n \leq N$.

With our definition, the operators $B_0,B_0^{\dagger}$ are not
assigned neither to the $bc$ sector nor to the $\tilde{b}\tilde{c}$
sector. On a generic state, the operator
$B^{\dagger}_0B_0$ takes the values $\b =0,1$ depending on whether the
level $n=0$ is empty or filled. While it takes no effort to keep $\b$
generic in the calculations, this is not really necessary, since the
configurations in which the level $n=0$ is empty have  a
Casimir $O(N^2)$ and therefore do not contribute in perturbation
theory, as we see from \eq{ZYM}.  In this section we limit ourselves
to the perturbative equivalence, and 
we can therefore restrict to the case $B^{\dagger}_0B_0=1$.

With these definitions,
\be\label{comm}
\{b_n,c_m\}=\d_{n+m}
\ee
and
\bees
c_n\fvac =0 &\hspace{1cm}&  n> 0\, ,\\
b_n\fvac =0 &\hspace{1cm}&  n \geq   0\, .
\ees
We now introduce an
auxiliary complex variable $z$ and we arrange 
$b_n,c_n$ into the modes of two holomorphic fields $b(z),c(z)$:
\bees
b(z)&\equiv& \sum_{n=-n_F}^{n_F}\frac{b_n}{z^{n+1}}\label{bz}\, ,\\
c(z)&\equiv& \sum_{n=-n_F}^{n_F}\frac{c_n}{z^{n}}\, .\label{cz}
\ees
Eqs.~(\ref{comm}) to (\ref{cz}) defines  a $bc$ theory with
$\l =1$ (see e.g. ref.~\cite{Pol}, sect.~2.7), with a cutoff at $|n|= n_F$.
The Fermi vacuum $\fvac$ corresponds, in the notation of
ref.~\cite{Pol}, to the vacuum state $|\downarrow\rangle$ of the $bc$
theory. Similarly, for  the modes $\tilde{b}_n,
\tilde{c}_n$ it follows from the definition that
\bees
\tilde{c}_n\fvac =0 &\hspace{1cm}&  n> 0\, ,\\
\tilde{b}_n\fvac =0 &\hspace{1cm}&  n \geq   0\, ,
\ees
and $\{\tilde{b}_n,\tilde{c}_m\} =\d_{n+m}$. It is  convenient to
arrange them into two antiholomorphic fields,
\bees
\tilde{b}(\bar{z})&\equiv& 
\sum_{n=-n_F}^{n_F}\frac{\tilde{b}_n}{\bar{z}^{n+1}}\, ,\\
\tilde{c}(\bar{z})&\equiv& 
\sum_{n=-n_F}^{n_F}\frac{\tilde{c}_n}{\bar{z}^{n}}
\, .
\ees
The fields $b(z), c(z)$ (and similarly for $\tilde{b},\tilde{c}$)
are just useful bookkeeping devices for
assembling together the modes $b_n, c_n$, and
there is nothing special in the choice $\l =1$. We could as well
assemble them into a $bc$ theory with $\l$ generic,
\be\label{lambda-generic}
b(z)\equiv \sum_{n=-n_F}^{n_F}\frac{b_n}{ z^{n+\l }}\, ,\hspace{1cm}
c(z)\equiv \sum_{n=-n_F}^{n_F}\frac{c_n}{z^{n+1-\l }}\, 
\ee
(and similarly for the $\tilde{b}\tilde{c}$ theory). However, 
the calculation of the bosonized form of the YM hamiltonian that we
will perform below turns out to be slightly simpler
when $\l =1$, so we will restrict to this choice. 
In  appendix~A we will check that the same final
result  for the string hamiltonian is obtained for $\l$ arbitrary.

A point to be kept in mind is that our $bc$ and $\bt\ct$
theories depend on $N$ 
through the cutoff, $|n|\leq n_F$. Furthermore, the two theories
are coupled by the
constraint $\sum_n \bnd B_n =N$, which can be rewritten as
\be
N=B^{\dagger}_0B_0 +\sum_{n=1}^N \bnd B_n +\sum_{n=-N}^{-1} \bnd
B_n=
1+ \sum_{n=-n_F}^{n_F} (c_{-n}b_n +\tilde{c}_{-n}\tilde{b}_n) \, ,
\label{vin1}
\ee
where we used the fact that $B^{\dagger}_0B_0=1$ on perturbative states.
We now
define $: (...) :$ as the normal ordering {\em with respect to}
$\fvac$, i.e. we anticommute 
the operators $b_n,c_m$ and $\tilde{b}_n,\tilde{c}_m$
until all destructors with respect to $\fvac$ (i.e.  
$c_n, \tilde{c}_n$  with  $n> 0$ and
$b_n, \tilde{b}_n$  with  $n\geq 0$)  are to the right. Of course,
this is different from the normal ordering with respect to
$\vac$. Then in \eq{vin1} the normal ordering exchanges all terms with 
$n=-n_F,\ldots ,-1$ both in $c_{-n}b_n$ and in
$\tilde{c}_{-n}\tilde{b}_n$,  and therefore
\eq{vin1} can be written as
\be
N = 1 + 2n_F + \sum_n :\, c_{-n}b_n +\tilde{c}_{-n}\tilde{b}_n\, :\, ,
\ee
and, since $n_F=(N-1)/2$, we get
\be\label{vin}
\sum_n :\, c_{-n}b_n +\tilde{c}_{-n}\tilde{b}_n\, :\,\, =0\, .
\ee

\subsection{Bosonization}

The bosonization of the $bc$ theory with $\l =1$ is  known to be
given by the linear dilaton theory~\cite{Pol}. However we have seen
that, at finite $N$, YM$_2$ is not exactly given by the product of
a $bc$ theory and a $\tilde{b}\tilde{c}$
theory, but there is also an
$N$-dependence which enters  through the cutoff on the mode number;
furthermore the $bc$ and $\tilde{b}\tilde{c}$ theories are coupled  through the
constraint (\ref{vin}).

As far as the cutoff is concerned, however,
we can see that if in \eqs{bz}{cz} we send the cutoff to
infinity, writing
\be
b(z)= \sum_{n=-\infty}^{\infty}\frac{b_n}{z^{n+1}}\, ,\hspace{1cm}
c(z)= \sum_{n=-\infty}^{\infty}\frac{c_n}{z^{n}}\, ,
\ee
the error that we are doing is exponentially small in $N$, and
therefore is irrelevant in the $1/N$ expansion. In fact, the fermionic
configurations in which some of the states with $|n|> n_F$
(where $n$ is the index of $b_n,c_n$,
i.e. it measures the excitation above the Fermi surface)
are occupied
correspond, through \eq{Young},
to Young diagrams with lines longer than $n_F$. From 
eqs. (\ref{C2}) to (\ref{Ctilde})
we see that the quadratic Casimir of these diagrams
are $O(N^2)$ and therefore, from \eq{ZYM}, the contribution of these
fermionic configurations to the partition function is 
$O(\exp \{-c \l AN\})$, with $c$ some positive constant.
These ``long'' Young diagram give therefore contributions that are
non-perturbative in the $1/N$ expansion. These will be the subject of
sect.~\ref{sect4}. In this section we limit ourselves to perturbation
theory. This means that, in bosonizing the $bc$ theory,
we can use the results valid in the infinity
cutoff limit, and  set to zero all modes $b_n,c_n$ with
$|n|>n_F$.  

The $bc$ theory can then  be bosonized using the standard formulas,
in terms of a holomorphic
field $X_L(z)$ (see e.g.~\cite{Pol}, sect.~10.3),
\be\label{bos1}
b=:e^{iX_L}:\, ,\hspace{5mm} c=:e^{-iX_L}:\, ,\hspace{5mm}
: bc : =i\pa X_L\, .
\ee
The normal ordering in this standard formula is just the normal
ordering with respect to $\fvac$ that we have used above (for
a $bc$ theory with $\l \neq
1$ this is actually not true, as discussed in appendix~A, and one
must be more careful).
Defining the modes $\a_m$ of $X_L$ from
\be\label{modes}
\pa X_L =i\sum_m\frac{\a_m}{z^{m+1}}\, ,
\ee
\eq{bos1} gives
\be\label{bosa}
\a_m = \sum_{n=-\infty}^{\infty} : c_{m-n}b_n :=
\sum_{n=-n_F}^{n_F} : c_{m-n}b_n :\, .
\ee
We have used the fact that perturbatively we can set
$b_n =0$ for $|n|>n_F$, to restrict the sum
over $-n_F\leq n\leq n_F$. 
Furthermore we can also restrict $|m - n| \leq n_F$, that implies $-(N-1)
\leq m \leq (N - 1)$.

The  energy-momentum tensor of the $bc$
theory with $\l =1$ can be written in terms of  $X_L$
as~\cite{Pol} 
\be
:(\pa b)c : -\pa :bc: = -\frac{1}{2} :\pa X_L\pa X_L : -
\frac{i}{2}\pa^2X_L
\ee
The right-hand side is the energy-momentum tensor of a linear dilaton
CFT. In terms of the Virasoro operators, we have
$L_m^{(bc)}=L_m^{(X)}$, 
with
\be
L_m^{(bc)} = \sum_{n=-\infty}^{\infty} (m-n) :b_nc_{m-n}:\, ,
\ee
\be
L_m^{(X)}=\frac{1}{2}\( \sum_{n=-\infty}^{\infty} 
:\a_{m-n}\a_n : \) -\frac{1}{2} (m+1)\a_m\, .
\ee
In particular, for $m=0$ we have
\be\label{L0}
L_0=\sum_{n=-\infty}^{\infty} n :c_{-n}b_n:\,\, =
\frac{1}{2}\a_0^2-\frac{1}{2} \a_0+
 \sum_{n=1}^{\infty} :\a_{-n}\a_n :  
\, . 
\ee
The $\tilde{b}\tilde{c}$ theory is bosonized similarly, in terms of an
antiholomorphic  field $X_R(\bar{z})$, whose modes we denote by
$\tilde{\a}_m$, with
\be\label{bosat}
\tilde{\a}_m = \sum_{n=-n_F}^{n_F} : \tilde{c}_{m-n}\tilde{b}_n
:\, .
\ee
The constraint (\ref{vin}) that relates the $bc$ and
$\tilde{b}\tilde{c}$ theories now becomes simply
\be\label{vina}
\a_0+\tilde{\a}_0 =0\, ,
\ee
so it is  a constraint between the
holomorphic and antiholomorphic parts of $X=X_L+X_R$.

The winding number $w$
of $X$ is defined as usual,  $w=\a_0-\tilde{\a}_0$. 
Eq.~(\ref{vina}) then means that
\be\label{aaw}
\a_0 =\frac{w}{2}\, ,\hspace{1cm}\at_0 =-\frac{w}{2}\, .
\ee
Writing
\bees
\a_0 & = \sum_{n=-n_F}^{n_F} : c_{-n}b_n : 
=\( \sum_{n=-n_F}^{n_F} c_{-n}b_n \) -n_F
&=\( \sum_{n=1}^N \bnd B_n \) -n_F
\, ,\\
\tilde{\a}_0 & = \sum_{n=-n_F}^{n_F} : \tilde{c}_{-n}\tilde{b}_n : 
=\( \sum_{n=-n_F}^{n_F} \tilde{c}_{-n}\tilde{b}_n \) -n_F
&=\( \sum_{n=-N}^{-1} \bnd B_n \) -n_F
\, ,
\ees
we see that
\be\label{wfilled}
w = ({\rm filled \,\, levels\,\, with\,\,  } n>0) -
 ({\rm filled \,\, levels\,\, with\,\,  } n<0)\, .
\ee
For $SU(N)$, we have seen that
representations that differ by an overall shift of the $n_i$,
$n_i\ra n_i + b $, with $b$ integer, are equivalent. We can use this
freedom to set $w=0$ and therefore $\a_0=\at_0=0$. For $U(N)$ instead
this is not so, because the $U(1)$ generator $Q=\sum_in_i$ is not
invariant under the shift, and therefore we must keep $w$ generic.
We can further notice that, since the number of fermions is fixed to be $N$,
there can be at most 
$N$ filled fermionic modes with $n>0$, in which case there are none with
$n<0$ and $w$ reaches its maximum values $w=N$, while in the opposite
case all fermions have $n<0$ and $w$ reaches its minimum value,
$w=-N$. Therefore for $U(N)$
\be\label{w}
-N \leq w \leq N\, .
\ee
We now want to write the hamiltonian for the $U(N)$ YM theory,
as well as the
$U(1)$ charge, in terms of $\a_n,\tilde{\a}_n$. The $U(1)$ charge 
$Q=\sum_i n_i$ in the second quantization formalism is
\be
Q=\sum_{n=-N}^N n\bnd B_n\, .
\ee
We rewrite it as
\bees
&&Q=\sum_{n=1}^N n\bnd B_n +\sum_{n=-N}^{-1} n\bnd B_n=
\sum_{n=-n_F}^{n_F} (n_F+1+n) c_{-n}b_n 
- \sum_{n=-n_F}^{n_F} (n_F+1+n) \tilde{c}_{-n}\tilde{b}_n =\nonumber\\
 &&= (n_F+1)\sum_{n=-n_F}^{n_F} ( c_{-n}b_n -\tilde{c}_{-n}\tilde{b}_n)
+ \sum_{n=-n_F}^{n_F} n  ( c_{-n}b_n -\tilde{c}_{-n}\tilde{b}_n)=
\nonumber\\
 &&= (n_F+1) (\a_0-\at_0) + (L_0-\tilde{L}_0 ) =
 \( n_F+\frac{1}{2}\) w + \sum_{n=1}^{N-1}
( \a_{-n}\a_n - \at_{-n}\at_n )\, ,\label{charge}
\ees
where in the last line we have used \eqs{L0}{aaw}.
So we find\footnote{Observe 
  that when $B_0^{\dagger}B_0$ has eigenvalue one, which is the
  case that we have considered, $w$ is an even number (since we are
  considering $N$ odd) so \eq{Q} gives
  an integer result for $Q$, as it should. If we repeat the
  computation when $B_0^{\dagger}B_0$ has eigenvalue zero we find that 
  the term $Nw/2$ in \eq{Q} is replaced by  $(N+1)w/2$ and, since we are
  considering the case $N$ odd, this is again an integer.
  Our result for $Q$ disagrees
  with ref.~\cite{Dou2}, eq.~(2.25), where the $w$ dependence is
  written $Nw$ instead  of $Nw/2$.}
\be\label{Q}
Q= \frac{N}{2} w +\sum_{n=1}^{N-1} (\a_{-n}\a_n -
\tilde{\a}_{-n}\tilde{\a}_n )\, .
\ee
The dependence on $w$ can be easily understood noting that, under the
constant shift $n_i\ra n_i +b$, with $b$ an integer, $Q=\sum_in_i\ra
Q+Nb$. This is correctly reproduced by \eq{Q}, since under 
$n_i\ra n_i +b$ we have $w\ra w+2b$, as we see from \eq{wfilled}. 

We now perform the bosonization of the $U(N)$ hamiltonian.
The $U(N)$ hamiltonian in second quantization reads
\be\label{HU}
H_{U(N)}=\frac{e^2L}{2}\[ \sum_{n=-N}^N n^2 \bnd B_n - \frac{N}{12}(N^2-1)\]
= \frac{e^2L}{2}\,   \sum_{n=-N}^N n^2 \, : \bnd B_n \,\, :
\ee
since  $N(N^2-1)/12$ is just the normal ordering constant, see
\eq{E0}. We write
\bees
&& \sum_{n=-N}^N n^2  :\, \bnd B_n\, : = 
\sum_{n=-n_F}^{n_F}  
(n_F+1+n)^2  \, :\,  c_{-n}b_n + \tilde{c}_{-n}\tilde{b}_n \, : =\nonumber\\
&=& (n_F+1)^2 \sum_{n=-n_F}^{n_F}  :\, c_{-n}b_n +
\tilde{c}_{-n}\tilde{b}_n  :\,
+2 (n_F+1 ) \sum_{n=-n_F}^{n_F} n  :\, c_{-n}b_n +
\tilde{c}_{-n}\tilde{b}_n  :\, +\nonumber\\
&+&\sum_{n=-n_F}^{n_F}  n^2  :\, c_{-n}b_n +
\tilde{c}_{-n}\tilde{b}_n  :\, \, .\label{HUbc}
\ees
The first sum vanishes because of \eq{vin}.  The second sum
is just $L_0+\tilde{L}_0$, and it is
immediately written in terms of 
$\a_n ,\tilde{\a}_n$ using \eq{L0}. Therefore 
\be\label{H1}
H_{U(N)}=\frac{e^2L}{2}\[ (N+1) (L_0+\tilde{L}_0) +
\sum_{n=-n_F}^{n_F}  n^2 : c_{-n}b_n + \tilde{c}_{-n}\tilde{b}_n : 
\]\, .
\ee
The last term in \eq{H1} can be bosonized using the identity
\be
\oint\frac{dz}{2\pi i}\, z^2 :\pa c\pa b : \, \, =\, 
-\sum_n n^2 : c_{-n}b_n : - \sum_n n : c_{-n}b_n :\, ,
\ee
which is easily checked substituting the mode expansion of the $b,c$
fields into the left-hand side. Similarly
\be
\oint\frac{d\bar{z}}{2\pi i}\, \bar{z}^2 :\bar{\pa} \ct\bar{\pa} \bt :\,\, =
\, +\sum_m m^2 : \ct_{-m}\bt_m : + \sum_m m : \ct_{-m}\bt_m :\, .
\ee
Eq.~(\ref{H1}) then becomes
\bees\label{H2}
H_{U(N)}=\frac{e^2L}{2}\[ N (L_0+\tilde{L}_0)
-\oint\frac{dz}{2\pi i}\, z^2 :\pa c\pa b : 
+\oint\frac{d\bar{z}}{2\pi i}\, \bar{z}^2 :\bar{\pa} \ct\bar{\pa} \bt :
\]\, .
\ees
Computing the OPE $\pa b(z) \pa c (0)$ we can derive the relation
\be\label{OPE}
:\pa b \pa c :\,\, = \,\,
\frac{i}{3} : (\pa X_L)^3: +\frac{i}{6} :\pa^3X_L:\, .
\ee
Then
\bees
&&-\oint\frac{dz}{2\pi i}\, z^2 :\pa c\pa b : 
+\oint\frac{d\bar{z}}{2\pi i}\, \bar{z}^2 :\bar{\pa} \ct\bar{\pa} \bt
: =
\frac{i}{3}\[  \oint\frac{dz}{2\pi i}\, z^2 :(\pa X_L)^3 : 
   - \oint\frac{d\bar{z}}{2\pi i}\, \bar{z}^2 :(\bar{\pa} X_R)^3 : \]
   + \nonumber\\
&&+ \frac{i}{6}\[  \oint\frac{dz}{2\pi i}\, z^2 :\pa^3 X_L : 
   - \oint\frac{d\bar{z}}{2\pi i}\, \bar{z}^2 :\bar{\pa}^3 X_R : \]
\, .
\ees
Substituting the mode expansion for $X$ into the above expression, 
the second bracket gives
$(-1/3) (\a_0 +\at_0 )$, which vanishes because of the constraint
(\ref{vina}). The first bracket gives instead a term cubic in  $\a
,\at$, and we get
\be\label{H3}
H_{U(N)}=\frac{e^2L}{2}\[ N (L_0+\tilde{L}_0)
+\frac{1}{3}\sum_{mnp}\d_{m+n+p} 
:\a_m\a_n\a_p+\at_m\at_n\at_p :\]\, .
\ee
In $\sum_{mnp}$ we separate  from
the rest the terms where $\a_0, \at_0$ appear,
\bees
&&\frac{1}{3}\sum_{mnp}\d_{m+n+p} 
:\a_m\a_n\a_p+\at_m\at_n\at_p : =\nonumber\\
&&\sum_{m,n>0} :\a_m\a_n\a_{-m-n}+\at_m\at_n\at_{-m-n} : +
\sum_{m,n<0} :\a_m\a_n\a_{-m-n}+\at_m\at_n\at_{-m-n} : +\nonumber\\
&&+\a_0 \sum_{m\neq 0} :\a_m\a_{-m}: 
+\at_0 \sum_{m\neq 0} :\at_m\at_{-m}: 
+\frac{1}{3}(\a_0^3+\at_0^3)\, .
\ees
Using \eqs{L0}{aaw}, and introducing $\l =e^2N$, which is the coupling
to be held fixed in the $1/N$ expansion, 
our final result for the $U(N)$ hamiltonian
reads\footnote{Our result disagrees with eq.~(4.48) of
  ref.~\cite{CMR}, where the dependence on $w$ 
  has been lost.}
\bees\label{HUN}
H_{U(N)}&=&\frac{\l L}{2} \[  
  \frac{w^2}{4} + \sum_{n=1}^{N-1} (\a_{-n}\a_n +\at_{-n}\at_n ) 
+\frac{1}{N}  w \sum_{n=1}^{N-1} (\a_{-n}\a_n -\at_{-n}\at_n )
+\right. \\
&+& \left. 
\frac{1}{N}\( \sum_{m,n>0}+\sum_{m,n<0} \)
 :\a_m\a_n\a_{-m-n}+\at_m\at_n\at_{-m-n} : 
\] \, .\nonumber
\ees
The Hamiltonian for $SU(N)$ is obtained subtracting $(e^2L/2) Q^2/N$
from \eq{HUN}. Using \eq{Q} we  find
\bees\label{HSUN}
H_{SU(N)}&=&\frac{\l L}{2}  \[  
\sum_{n=1}^{N-1} (\a_{-n}\a_n +\at_{-n}\at_n ) -
\frac{1}{N^2} 
\( \sum_{n=1}^{N-1} (\a_{-n}\a_{n} -\at_{-n}\at_n )   \)^2 +
 \right. \\
&+&  \left.
\frac{1}{N} \(  \sum_{m,n>0} +\sum_{m,n<0} \)
:\a_m\a_n\a_{-m-n}+\at_m\at_n\at_{-m-n} : 
\right] 
\, . \nonumber
\ees
Observe that for $SU(N)$ the dependence on the winding number $w$
cancels, as it should, since we have seen that for $SU(N)$ we could
have set $w=0$ from the beginning. The cancellation is however a check
of the correctness of \eqs{HUN}{Q}.

Eqs.~(\ref{HUN}) and (\ref{HSUN})  prove
that, at least in perturbation theory in $1/N$, 
YM$_2$ with gauge group $U(N)$ or $SU(N)$
is equivalent to a string field
theory, described by a string 
field $X(z,\bar{z})$, and governed by an
Hamiltonian consisting of terms $O(1)$
and $O(1/N)$  (and, for $SU(N)$, a quartic term $O(1/N^2)$), describing the
creation and annihilation of strings. We have seen explicitly that, at
least perturbatively, 
eqs.~(\ref{HUN}) and (\ref{HSUN}) are {\em exact}, i.e. there are
no further terms suppressed by powers of $1/N$.
Eq.~(\ref{HSUN}) coincides with \eq{H} and correctly
reproduces the $1/N$ expansion of $SU(N)$ YM$_2$.

\section{The non-perturbative correspondence}\label{sect4}

\subsection{D-branes from YM$_2$}\label{sect4.1}

From \eq{ZYM} we see that Young diagrams with a quadratic Casimir 
$C_2=O(N^2)$ give contributions to $Z_{\rm YM}$ proportional to
$\exp\{-O( N)\}$; limiting ourselves for simplicity to a torus target
space (so that  $({\rm dim}\, R)^{2-2G}=1$ in \eq{ZYM}),
the structure of $Z_{\rm YM}$ is
\be\label{str}
Z_{\rm YM}=\[ O(1) + O(1/N^2) +\ldots \] + O(e^{-O( N)})
\ee
where the bracket represents the perturbative expansion discussed
above. 
If the string-YM correspondence holds even beyond perturbation theory
in $1/N$, the terms $e^{-O( N) }$ should
correspond to terms $e^{-O(1/g_s) }$ on the string theory side.
In the following, for
definiteness, we will consider the case of $SU(N)$.

An exact evaluation of the  contributions $e^{-O( N)}$
to $Z_{\rm YM}$ seems to be a quite formidable task. However, there is
a large class of diagrams that we are able to evaluate, and which will turn
out to give a rather interesting result. These are the Young diagrams
in which one or more lines have more than $N$ boxes and the remaining part of
the diagram has a  number of boxes $O(1)$, see
figs.~\ref{fig2} and \ref{fig3}.
\DOUBLEFIGURE[h]{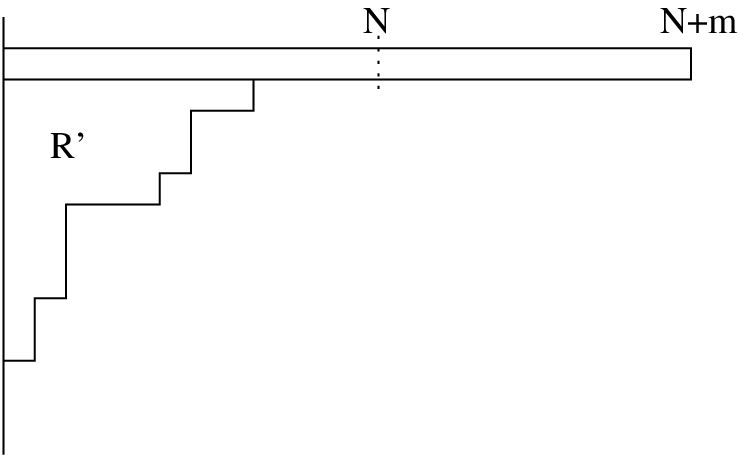, width=.4\textwidth}
{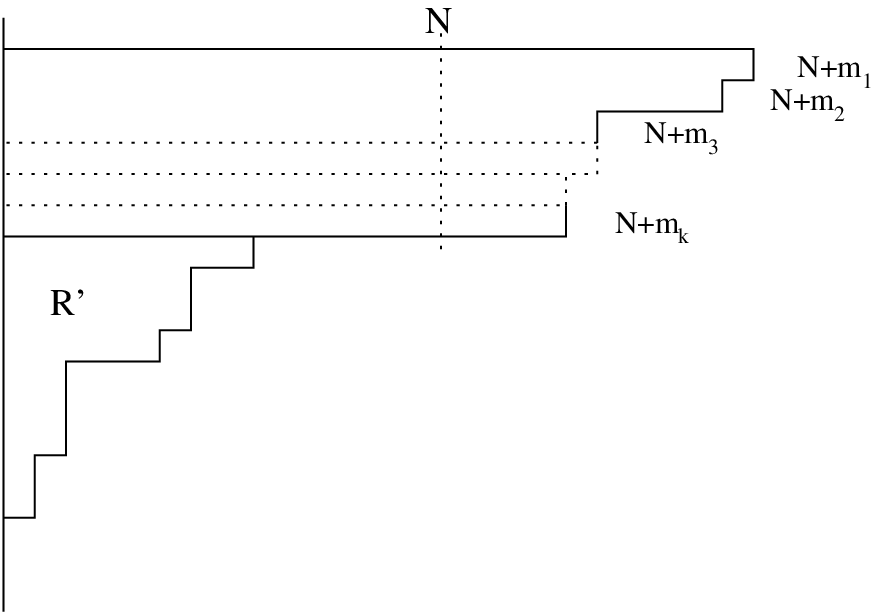, width=.4\textwidth}{\label{fig2} Diagram with the first
line longer than $N$.}{\label{fig3} Generic diagram with $k$ lines
with more than $N$ boxes.} 
Thus, we find useful to introduce a distinction
between ``bounded'' diagrams, defined as those diagrams in which all
lines have less than $N$ boxes, and ``long'' diagrams, i.e. those in
which at least the first line, and possibly more lines,
are longer or equal to  $N$.
In particular one can consider long diagrams with $k$
long lines, and long diagrams with $N-k$ long lines: the contribution
to the partition function of the two groups of diagrams is the same,
since each diagram of the second group has the same Casimir of a
complementary diagram of the first group, where the correspondence is
the one shown in fig.~\ref{fig6}.
\EPSFIGURE[h]{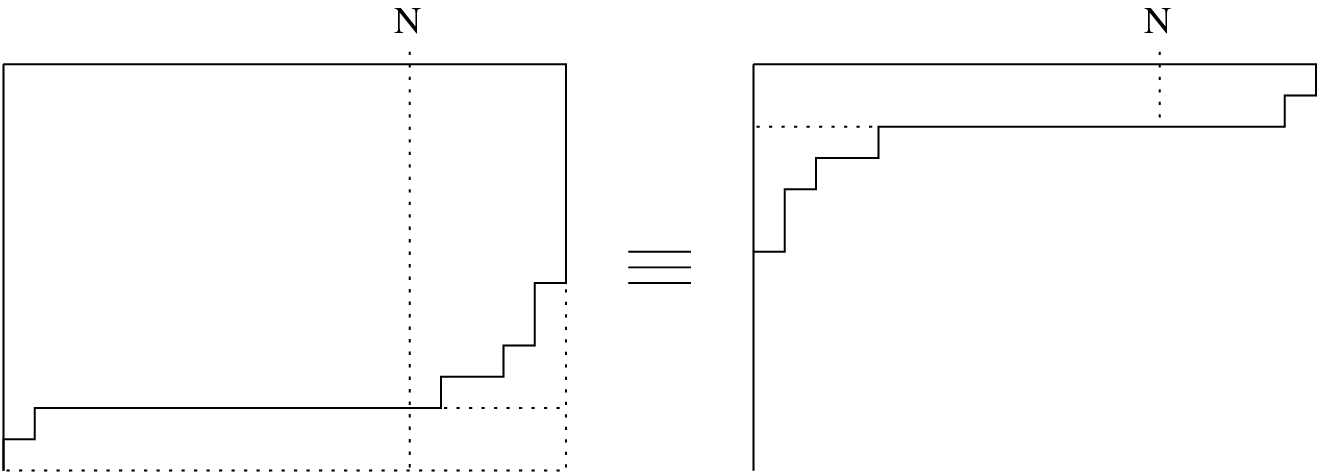, width=.43\textwidth}{\label{fig6} Diagram with $O(N)$ long lines and its complementary, both with the same Casimir value.} 
From the expression
of the Casimir, \eq{C2SUN}, one sees immediately that long
diagrams are interesting candidates for
non-perturbative contributions, since they have $C_2
=O(N^2)$ and therefore their contributions to $\exp\{ (-\l A/2N)
C_2\}$ is $\exp\{ -O(N)\}$. 
However,  they certainly do not exhaust the class of
all Young diagrams with $C_2=O(N^2)$, since in general 
diagrams  with $O(N)$ boxes in the bounded part $R'$ can
have $C_2=O(N^2)$, independently of whether they have long lines or
not.  While 
the contribution of diagrams with $O(N)$ boxes in the bounded part
$R'$ 
is difficult to evaluate, the contribution of long diagrams with 
$O(1)$ boxes in $R'$ can be evaluated as follows.

Consider first a long diagram as the one shown in fig.~\ref{fig2}, with
the first line of length $N+m$, with $m=0,1,\ldots \infty$ generic,
and $O(1)$ boxes in the remaining part. 
Since we are considering $SU(N)$, there are at most $N-1$ lines in total.
Eliminating the first line, we
are left with a Young diagram corresponding to a generic
representation $R'$ of a chiral sector (in the sense of \cite{GT})
of $SU(N-1)$. Let again $h_i$ be the
number of boxes in the $i$-th line, $n=\sum_{i=1}^{N-1}h_i$ the total
number of boxes in the diagram $R$, and let
$n'=\sum_{i=2}^{N-1}h_i$
be the total number of boxes in $R'$. Simple algebra shows that
\be
C_2(R)
=C_2(R')+m\(3N-3-\frac{2 n'}{N}\)+ m^2 \(1 -\frac{1}{N}\) + 2N^2
-2N-4n'
\, .\label{CRR}
\ee
Since $n'=O(1)$, the leading terms in $C_2(R)$ are:
\be\label{C'}
C_2(R)\simeq C_2(R')+ 2N^2 +N(3m-2)+m^2
\, .
\ee
The great simplification in \eq{C'} is that $n'$ does not appear
explicitly and all dependence on
$R'$ is through  $C_2(R')$. This  
allows to factorize the contributions
of the subdiagram $R'$. In fact,
summing over all $m=0,\ldots ,\infty$ and over all representations
$R'$ with $n'=O(1)$, and defining
\be
a\equiv \frac{\l A}{2}\, ,
\ee
we find that the contribution to $Z_{\rm YM}$ 
of this class of diagrams is
\bees
&&\sum_{R'}\sum_{m=0}^{\infty}
e^{ -\frac{a}{N}\[ C_2(R') +2N^2 +N(3m-2)+m^2 \]}
\( 1+O\(\frac{1}{N}\) \)=\nonumber\\
&=&e^{2a} e^{-2aN} \[ \sum_{R'}e^{ -\frac{a}{N} C_2(R')} 
 \sum_{m=0}^{\infty}e^{ -\frac{a}{N}(3mN+m^2 )} 
\( 1+O\(\frac{1}{N}\) \)\]=\nonumber\\
&=&  e^{-2aN} \,\, \[\(  Z^{\rm chir}_{SU(N-1)}\)
\frac{e^{2a}}{1-e^{-3a}} +O\(\frac{1}{N}\)\]\, .\label{1l}
\ees
We have denoted by $Z^{\rm chir}_{SU(N-1)}$ the chiral partition
function of $SU(N-1)$ (see ref.~\cite{GT})\footnote{We should remark that
the factorization of 
the contribution of  the representations $R'$ takes place only at leading
order. The $O(1/N)$ corrections in \eq{1l} are not simply proportional
to $ Z^{\rm chir}_{SU(N-1)}$.}.

It is not difficult to extend this result to diagrams with $k>1$ long
lines, with $k\ll N$ (see fig.~\ref{fig3}). Let the length of the long lines be
$h_i=N+m_i$, with $m_1\geq m_2\geq \ldots \geq m_k$. As discussed
above,  an identical contribution comes from diagrams with $N-k$ long lines;
then, with the same approximations used in
\eq{C'}, we find
\be\label{C'k}
C_2(R)\simeq C_2(R')+ 2k N^2 +(3\sum_{i=1}^km_i-2k^2)N+\sum_{i=1}^km_i^2\, .
\ee
The resummation of all contributions with $k$ long lines, with $k\ll
N$, gives therefore
\bees
&&\sum_{R'}
\sum'_{m_1, \ldots m_k} 
e^{ -\frac{a}{N}\[ C_2(R') +2k N^2 +(3\sum_{i=1}^km_i-2k^2)N+\sum_im_i^2
\]}
\( 1+O\(\frac{1}{N}\) \)=\nonumber\\
&=&  e^{-2kaN}\,\, \[\( Z^{\rm chir}_{SU(N-k)}\)
e^{2k^2a}\prod_{m=1}^k \frac{1}{1-e^{-3am}} +O\(\frac{1}{N}\)\]\, .
\ees
(where in the first line
$\sum_{m_1, \ldots m_k}'$ runs over all $m_i=0,\ldots \infty$ with
the condition $m_1\geq m_2\geq\ldots\geq m_k$).
Thus, we have been able to resum a very large class of diagrams, and
the result is quite interesting: the resummation of all
diagrams with $k$ ``long'' lines, i.e. with $k$ lines longer than $N$,
and with all possible chiral subdiagrams $R'$ with $O(1)$ boxes
gives a contribution proportional to
\be
 e^{-2kaN}\, , \hspace{2cm} k=1,2,\ldots 
\ee
An equal contribution comes  from the resummation of the diagrams
with $N-k$ long lines.

Recalling that $N=1/g_s$, $a=\l A/2$ and that $\l$ is related to the
string tension of the string theory by $\l =1/(\pi\a' )$ (see the
discussion below \eq{log}), we see that these contributions are just
of the form $e^{-S_{D1}}$
with
\be\label{S1}
S_{D1}=\tau_1 kA\, ,
\ee
and
\be
\tau_1 =\frac{1}{\pi \a' g_s}\, .
\ee
Now, $\tau_1$ has exactly the form expected for the tension of a
D1-brane, modulo a numerical factor which depends on the specific
theory (for instance in type~IIB in 10 dimensions, $\tau_1=1/(2\pi \a'
g_s)$~\cite{Pol}). The dependence on the target space area is also 
what we would expect from $D1$-branes. Indeed, recall that the string
theory equivalent to YM$_2$ is quite peculiar because it describe a
string with no foldings~\cite{Gross}, i.e. a string whose world-sheet
area is an integer times the target space area. The integer then
counts the number of times that the string world-sheet covers the
target space. If this theory has $D1$-branes, it is therefore natural
to expect that they, too, have no foldings, and indeed the factor
$kA$ in \eq{S1} can be interpreted as the world-sheet area of a $D1$-brane
wrapping $k$ times, without foldings, over the target space.

Instead, it is clear that (independently of any approximation)
there is no contribution that could be
interpreted as $e^{-S_{D0}}$, with $S_{D0}$ the action of a
$D0$-brane. In fact, $S_{D0}$ would rather be proportional to the
length of the 
world-line of the $D0$-brane. However, $Z_{YM}$ is a function only of
the area of the target space, which has no relation to the 
world-line length
and therefore  such terms are absent.

Thus, we have a non-perturbative structure in which 
terms that allow an interpretation as $Dp$-brane with
$p$ odd (i.e. $p=1$ because we are in two dimensions) are present,
while with $p$ even (i.e. $p=0$)
they are absent. This is the typical structure of 
a type~B string theory. Since we have no spacetime supersymmetry, it
is quite natural to identify the theory with a sort of type~0B
string theory. 

Finally, it is interesting to recall that perturbation theory is
really  an expansion in $1/N^2$, i.e. in $g_s^2$, rather than in $1/N$. It
was in fact observed by Gross \cite{Gross} 
that the terms with odd powers in $1/N$
are zero because of a cancellation between a Young diagram $R$ and its 
conjugate $\bar{R}$ which has its rows and columns
interchanged. However, when we consider ``long'' diagrams, 
i.e. diagrams with lines longer than $N$, the
conjugate diagram does not exist, because we cannot have 
columns with  more than
$N$ boxes. So the cancellation does not take place, and in the
non-perturbative sectors the corrections have the form
\be
e^{-k\frac{2a}{g_s}} \( 1+O(g_s) \)\, ,
\ee
rather than $e^{-k\frac{2a}{g_s}} (1+O(g_s^2))$.

\subsection{The stringy exclusion principle}\label{exclusion}

In the previous section we have understood the effect of ``long''
Young diagrams: we have seen that the Young
diagrams with $k$ (or $N - k$) lines longer than $N$ give the
contribution that, in 
string theory, would be expected from a $D1$-brane
wrapping $k$ times over the target space. We
now turn our attention to the non-perturbative effects in
$Z^{\rm bounded}_{SU(N)}$, again limiting ourselves to the torus. 
Following ref.~\cite{Gross}, we consider the contribution of
the (bounded) diagrams in which the total number of boxes $n$ is $O(1)$,
rather than $O(N)$.\footnote{In the language of sect.~\ref{sect2}
this means  that we are restricting to  excitations around the
Fermi surface at $+n_F$. A similar contribution comes from the
excitations around $-n_F$. For the torus, at leading order, this just
results in an overall factor of 2, which is not important for our
purposes. In the notation of \cite{GT,CMR}, we are restricting to one
chiral sector.} Then in the Casimir (\ref{C2SUN}) the term
$\tilde{C}$ is $O(1)$ while  $n^2/N=O(1/N)$, so they can both
be neglected compared
to $nN$. Therefore in this approximation~\cite{Gross}
\be
Z_{\rm YM}^{G=1}\simeq \sum_{\{h_i\}}\, e^{-a\sum_{i=1}^{N-1} h_i}\, ,
\ee
where as usual  $h_i$ denote the length of the $i$-th row and
therefore $\sum_{\{h_i\}}$ runs over 
the domain $h_1\geq h_2\geq\ldots\geq
h_{N-1}\geq 0$. 
The sum is performed~\cite{Gross}
introducing $k_1=h_1-h_2, k_2=h_2-h_3,\ldots ,k_{N-2}=h_{N-2}-h_{N-1},
k_{N-1}=h_{N-1}$. Then $\sum_ih_i=\sum_j jk_j$ and
\be
Z_{\rm YM}^{G=1}\simeq \sum_{k_1=0}^{\infty}\ldots\sum_{k_{N-1}=0}^{\infty} 
\, e^{-a\sum_{j=1}^{N-1} jk_j}=
\prod_{m_1=1}^{N-1} \( \sum_{k=0}^{\infty} e^{-am_1 k}\)\, .
\ee
Then
\be
Z_{\rm YM}^{G=1}\simeq 
\prod_{m_1=1}^{N-1} \frac{1}{1-e^{-am_1}}\, .
\ee
From the non-perturbative point of view,
the interesting aspect of this result is 
that the product over $m$ runs only from $m=1$ to $m=N-1$, rather than
up to $m=\infty$. The reason, of course, is that there are only $N-1$
variables $k_j$ because the Young diagrams of $SU(N)$ have at most
$N-1$ lines. Taking the logarithm and expanding it, 
\be\label{m2}
\ln Z_{\rm YM}^{G=1}\simeq -\sum_{m_1=1}^{N-1}\ln (1-e^{-am_1})=
\sum_{m_1=1}^{N-1}\sum_{m_2=1}^{\infty}\frac{1}{m_2} e^{-am_1m_2}\, .
\ee
Inserting $1=\sum_{n=1}^{\infty}\d_{m_1m_2,n}$, 
\be\label{Zc}
\ln Z_{\rm YM}^{G=1}\simeq \sum_{n=1}^{\infty} 
c(n) e^{-an}\, ,
\ee
with 
\be\label{cn}
c(n)=\sum_{m_1=1}^{N-1}\sum_{m_2=1}^{\infty}
\frac{1}{m_2}\d_{m_1m_2,n}\, .
\ee 
Eqs.~(\ref{Zc}) and (\ref{cn}) show clearly the geometric
interpretation in terms of a theory of maps. In fact,
$e^{-an}=\exp\{ -(\l /2)An\}$ 
is just the factor expected from a string
without foldings that wraps $n$ times around the target space,
with string tension $1/(2\pi\a')=\l /2$, while it is possible to show
that  $c(n)$ is just
the number of coverings of the torus by a torus with $n$
sheets~\cite{Gross,Gross2}. Thus we have a mapping from a world-sheet to
a target space, i.e. a string, and
in this interpretation $m_1,m_2$ are the
number of times that the string world-sheet winds around the two cycles of the
torus.

The surprise, in \eq{cn}, is that the winding over one of the cycles,
$m_1$, is limited by $N-1$ for $SU(N)$ (or by $N$ if we repeat the
calculation for $U(N)$). So, first of all, there is an asymmetry
between $m_1$ and $m_2$, which instead ranges from $1$ to
$\infty$. Technically this came out because $m_1$ and $m_2$ have a
very different origin: $m_1$ labels the variables $k_j$
and therefore the lines in a Young diagram, and then it
cannot exceed $N-1$. Instead $m_2$ appeared from the Taylor expansion
of the logarithm in \eq{m2}. However, it is clear that this asymmetry
must be an artefact of our approximations, i.e. of restricting to the
class of Young diagrams such that $\tilde{C}$ can be neglected, 
and if one would be able to
compute exactly the non-perturbative contributions the symmetry should
be restored. Our expectation is that both cycles will then be limited
by $N-1$. Therefore, the number of times that the string winds on the
target space torus is limited by a value $N-1$ for $SU(N)$ or
$N=1/g_s$ for $U(N)$. This is clearly a
non-perturbative limitation, and it is very similar to the
stringy exclusion principle found by Maldacena and
Strominger~\cite{MaldaStro} in the context of AdS$_3$.

\subsection{Numerical investigation of  the
non-perturbative phase structure}

Given the difficulty of an exact analytical investigation of 
the non-perturbative contributions, one might consider a numerical
study. Actually, the ``long'' diagrams discussed in
sect.~\ref{sect4.1} would be difficult to study numerically, because
even for fixed $N$ there is an infinite number of them; however, we
have shown that these diagrams can be resummed and can be
well understood
analytically. A complete analytic understanding is instead
more difficult for
the ``bounded'' diagrams but since, at fixed $N$, there is a finite
number of them, one could try to compute their effect numerically. 
In particular,
one might try a strategy borrowed from lattice gauge theory
simulations: evaluate the partition function (\ref{ZYM}) numerically, 
restricting the sum to the bounded diagrams; subtract the
perturbative contribution, evaluated to a sufficiently large order,
chosen such that, numerically, the exponential terms can be extracted
by a fit against $N$.
Furthermore, the perturbative contribution to the torus partition
function have already been computed explicitly to very large order
in ref.~\cite{Rudd}. 

This strategy however meets an instructive problem. Fig.~\ref{fig4}
shows the ``bounded'' partition function of the torus, evaluated
numerically for 
different values of $N$ and $a$, and compares it with the perturbative
expansion of ref.~\cite{Rudd}, pushed up to 6th order.\footnote{Of
  course  $Z_{\rm YM}^{\rm bounded}$ and the full partition
  function $Z_{\rm YM}$ differ only by the contribution of ``long''
  diagrams, so they have the same perturbative expansion.} As we
expect, for sufficiently large $N$ the two coincide, and there is 
a critical value $N_c$ below which they start to diverge; 
actually, at $N<N_c$
even the qualitative behaviour of $Z_{YM}^{\rm bounded}$ 
has nothing to do with its
perturbative expansion.

Numerically, we have found that the critical value $N_c$ is  a
decreasing function of $a$, roughly given 
by $aN_c(a)\sim \g$, with $\g$ a numerical constant.\footnote{The numerical
  value of $\g$ depends of course on the precise definition of
  $N_c$. For instance, if $N_c$ is defined as the point where the
6th order perturbative series and the numerical result differ by
$5\%$, then  $\g=O(20)$. Also, the precise functional form of $N_c(a)$
is not exactly $\sim 1/a$. However, the only important point for us
is simply that there are two qualitatively different regions separated
by a curve $N_c(a)$.}
 This means
that the perturbative expansion is a good approximation only for
$aN\gg \g$; 
\EPSFIGURE[h]{fig4.eps, width=.6\textwidth}{\label{fig4}  Numerical 
evaluation of
$Z_N^{\rm bounded}$. Dashed lines are the plots of the perturbative series. } 
Taking as a typical reference value the non-perturbative contributions
$\sim e^{-2aN}$ found from long diagrams,
we see that, when the perturbative expansion starts to be 
in rough agreement with the exact result,  a term of this type
would be already
suppressed at least by a factor
$e^{-2aN}\sim e^{-2\g }=O(10^{-18})$ compared to
the perturbative term which is $O(1)$, and it is therefore numerically
invisible. 

However, the fact that $aN_c(a)\sim\g$ is of some interest in
itself. It means that in the plane $(g_s, a)$ there is a
non-trivial phase structure. When $g_s\ll a/\g$, perturbative string
theory is a good approximation to the full theory
(at least if at the same time $g_s\leq 1/2$, because $g_s=1/N$ 
and $N\geq 2$). Instead, when $g_s\sim a/\g$
we enter  into a qualitatively different regime,  
as we see from fig.~\ref{fig4},
where the perturbative expansion is of no use and strong coupling
effects are dominant. 

If we take the limit $a\ra 0$ at fixed $g_s$, 
we   always end up in this strong coupling domain, for all non-zero values
of $g_s$. The limit $a\ra 0$  has been studied
in ref.~\cite{CMR}, where  it is found 
that YM$_2$ becomes a topological  string theory. This therefore
clarifies the nature of the theory in the strong coupling phase 
$g_s\gsim a/\g$. On the other hand, this also means that from the limit
$a\ra 0$ we cannot learn anything about the perturbative string
theory, since the two regimes are qualitatively different.

\section{Conclusions}

We have examined various aspects of the string/YM correspondence in two
dimensions. At the perturbative level we have shown how,
from the bosonization of
the fermionic formulation of YM$_2$, one can derive rigorously the string
field theory hamiltonian which reproduces the full $1/N$ expansion of
the theory. At the non-perturbative level, we have found that the
YM$_2$ partition function reproduces a number of non-perturbative
effects which should be expected in the corresponding string
theory. In particular, we have identified representations of $SU(N)$
that would correspond to $D1$-branes in the string formulation,
while terms that could be identified with $D0$-branes are absent;
this suggests that the correspondence holds even non-perturbatively,
and that the non-perturbative structure is typical of a type~0B string
theory. 

We conclude  with some  conjectural remarks.
If the interpretation in terms of some form of type~0B theory 
on the cylinder is
correct, it is natural to ask what happens if we perform a T-duality
transformation along the compact spatial direction 
of the cylinder, and it is natural to expect to get a
type~0A string theory; the $D1$-branes would then become $D0$-branes.
Such a string theory would not have a direct relation with a
two-dimensional YM, since we have seen that in YM$_2$ the partition
function depends only on the area of the target space, and cannot
account for the effect of $D0$-branes.

However, a type~A theory, and D0-branes, could be the signal of 
the non-perturbative
opening up of a third dimension, with size $R_3\sim g_s\a'^{1/2}$.
Of course, since we have no space-time supersymmetry, the possibility
of the opening of a third dimension, and correspondingly the existence
of a three-dimensional M-theory, should be taken with the same caveats
that hold for the  bosonic string in 26 dimensions. Even in that
case, however, there are arguments suggesting the existence of a
27-dimensional M-theory~\cite{HS}.

If these conjectures are correct, there should be a 3-dimensional
M-theory which reduces to a two-dimensional string theory 
at weak coupling, when the third dimension becomes unaccessible. It is
quite tempting to conjecture that such an M-theory could be  a
Chern-Simons (CS) theory on a suitable manifold with a boundary. 
This is suggested by the well known fact that a CS theory on a
three-dimensional manifold with a boundary induces a 
current algebra on the
boundary~\cite{W1}, and indeed CS theory can be used to produce 
in this way all rational
CFT~\cite{MS}. Furthermore, it is possible to  construct 
string theories,  which have the peculiarity that
the matter and ghost sectors do not decouple, which have the target space
interpretation of a CS theory~\cite{W2}.

\vspace{2cm}

\appendix

\section{Bosonization and string hamiltonian for $\l$ generic}

In this appendix we repeat the calculations that led to the
string hamiltonian starting from \eq{lambda-generic} with $\l$
generic. 
This is an useful check of the correctness of the result, and will
reveal some small subtlety  in the computation, especially concerning
the relevant definition of normal ordering.

The formulas for the bosonization of the $b c$ ($\tilde{b} \tilde{c}$) 
theory are the standard ones used in sect.~\ref{sect3}:
\be\label{bos-gener}
b = : e^{iX_L} :_c\, ,\hspace{5mm} c = : e^{-iX_L} :_c\, ,\hspace{5mm}
: bc :_c =i\pa X_L\, .
\ee 
(and similar ones for the $\tilde{b} \tilde{c}$ fields, with an
antiholomorphic bosonic field $X_R(\overline{z})$). However,
it is important to observe 
 that the normal ordering in this relations is the
conformal one, that in the $b c$ theory is related to the
annihilation-creation one by (see e.g.~\cite{Pol}, chapt.~2):
\be\label{n-o-rel-pol}
:b(z) c(z'):_c \, = \, :b(z) c(z'): +\, \frac{(z/z')^{1 - \lambda} -
1}{z - z'} 
\ee  
from which one can derive:
\be\label{n-o-rel}
:b(z) c(z):_c \, = \, :b(z) c(z): +\, \frac{1 - \lambda}{z}
\ee 
We see that for $\l = 1$ they are equal; therefore in this special
case we could neglect the distinction between the two.

For the bosonic theory instead the normal ordering of
annihilation-creation (with respect to the standard vacuum) is identical
to the conformal one. 

Developing the $X$ field in modes as in the $\l = 1$ case
(\eq{modes}), using \eq{bos-gener} (and the analogous ones for the
$\tilde{b} \tilde{c}$ fields) and the relation (\ref{n-o-rel}) between
the normal orderings, we obtain:
\bea
\alpha_m & = & \sum_{n = - n_F}^{n_F} :c_{m - n} b_n: + (\lambda - 1)
\delta_{m, 0}\label{rel-modi}\\
\tilde{\alpha}_m & = & \sum_{n = - n_F}^{n_F} :\tilde{c}_{m - n}
\tilde{b}_n: + (\lambda - 1) \delta_{m, 0}\nonumber
\eea    
In particular the constraint (\ref{vin}) becomes:
\be\label{vinc-bos}
\alpha_0 + \tilde{\alpha}_0 = 2 (\lambda - 1)
\ee
In the general case the Virasoro generators of the $b c$ theory are:
\be
L_m^{(bc)} = \sum_n\, (m \lambda - n) :b_n c_{m -n}: +\, \frac{\lambda(1 -
\lambda)}{2} \delta_{m, 0}
\ee
In particular:
\be\label{L0-bc}
L_0^{(bc)} = \sum_n n :c_{-n} b_n: + \,\frac{\lambda(1 - \lambda)}{2} 
\ee
The Virasoro generators of the bosonic theory are:
\be
L_m^{(X)}=\frac{1}{2} \sum_{n=-\infty}^{\infty} :\a_{m-n}\a_n :  
-\(\l - \frac{1}{2}\) (m+1)\a_m\, .
\ee
and in particular:
\be\label{L0-bos}
L_0^{(X)}  =  \frac{1}{2} \alpha_0^2 + \sum_{n=1}^{\infty}
:\a_{-n}\a_n : - \(\lambda - \frac{1}{2}\) \alpha_0
\ee
The antiholomorphic field obeys similar formulas.
To obtain the hamiltonian (for $U(N)$ and $SU(N)$) we can easily
generalize the calculations of sect.~\ref{sect3}.

For what concern the $U(1)$ charge, following the same steps of
\eq{charge}, we obtain:
\bea
Q & = & (n_F+1) (\a_0-\at_0) + (L_0-\tilde{L}_0) = 
(n_F+1) (\a_0-\at_0) +
(\a_0-\at_0)(\lambda - 1) + \nonumber\\
& - & (\l - \frac{1}{2})(\a_0-\at_0)
+ \sum_{n=1}^{N - 1}( \a_{-n}\a_n - \at_{-n}\at_n )
\eea 
where we have used \eq{L0-bos} and \eq{vinc-bos}. So we finally obtain
the same result of the $\l = 1$ case, as we expected:
\be
Q= \frac{N}{2} w +\sum_{n=1}^{N - 1} (\a_{-n}\a_n -
\tilde{\a}_{-n}\tilde{\a}_n )\, .
\ee

The $U(N)$ hamiltonian, given by \eq{HUbc}, can be written as:
\be
H_{U(N)} = \frac{e^2}{2}\, L \[\sum_{n = -n_F}^{n_F}  n^2 (:c_{-n}
b_n: + :\tilde{c}_{-n} 
\tilde{b}_n:) + (N + 1) (L_0 + \tilde{L}_0) - (N + 1) \lambda (1 -
\lambda)\]
\ee    
For $\l$ generic we must use the relation:
\bea
\oint \frac{d z}{2 \pi i} z^2 :\partial c \partial b: & = & - \sum_n
n^2 :c_{-n} b_n: + (1 - 2 \lambda) \sum_n n :c_{-n} b_n: +
\nonumber\\
& & + \lambda (1 - \lambda) \sum_n :c_{-n} b_n:
\eea
and the similar one for tilded fields:
\bea
\oint \frac{d \overline{z}}{2 \pi i} \overline{z}^2
:\overline{\partial} \tilde{c} 
\overline{\partial} \tilde{b}: 
& = & + \sum_n 
n^2 :\tilde{c}_{-n} \tilde{b}_n: - (1 - 2 \lambda) \sum_n n
:\tilde{c}_{-n} \tilde{b}_n: + 
\nonumber\\
& & - \lambda (1 - \lambda) \sum_n :\tilde{c}_{-n} \tilde{b}_n:
\eea
Thus the hamiltonian becomes:
\bea
H_{U(N)} & = & \frac{e^2}{2}\, L \[- \oint \frac{d z}{2 \pi i} z^2
:\partial c \partial b: + \oint \frac{d \overline{z}}{2 \pi i}
\overline{z}^2 
:\overline{\partial} \tilde{c} \overline{\partial}
\tilde{b}: + \right.\nonumber\\
& & \left. + (N + 2 - 2 \lambda)(L_0 + \tilde{L}_0) 
- (N + 2 - 2 \lambda) \lambda (1 - \lambda)\phantom{\oint}\!\!\!\!\!\!\]
\eea
(where we have used \eq{vin} and \eq{L0-bc}).
Using \eq{n-o-rel-pol} we can derive the relation:
\be\label{relaz-no-der}
:\partial c(z) \partial b(z): \, = \, :\partial c(z) \partial b(z):_c +\,
\frac{\lambda^3 - 3 \lambda^2 + 2 \lambda}{3 z^3}
\ee
(and analogously for the $\tilde{b} \tilde{c}$ fields).

Using \eq{relaz-no-der}, then \eq{OPE} and finally the mode expansion
of the bosonic field, we obtain:
\bea
H_{U(N)} & = & \frac{e^2}{2}\, L \[\( \sum_{m,n>0}+\sum_{m,n<0} \)
 :\a_m\a_n\a_{-m-n}+\at_m\at_n\at_{-m-n} : + 2 \alpha_0 \sum_{n=1}^{N
- 1} \a_{-n}\a_{n} +   
\right.\nonumber\\
& & +  2 \tilde{\alpha}_0
\sum_{n=1}^{N - 1} \at_{-n}\at_n  
+ \frac{1}{3} \alpha_0^3 + \frac{1}{3} \tilde{\alpha}_0^3 -
\frac{1}{3} (\alpha_0 + \tilde{\alpha}_0) + (N + 2 - 2 \lambda) (L_0 +
\tilde{L}_0) + \nonumber\\
& & \left. - (N + 2 - 2 \lambda) \lambda (1 - \lambda) - \frac{2}{3}
(\lambda^3 - 3 \lambda^2 + 2 \lambda)\right] 
\eea
which using \eq{L0-bos} and \eq{vinc-bos} becomes:
\bees
H_{U(N)}&=&\frac{(e^2N) L}{2} \[
  \frac{w^2}{4} + \sum_{n=1}^{N - 1} (\a_{-n}\a_n +\at_{-n}\at_n ) 
+\frac{1}{N}  w \sum_{n=1}^{N - 1} (\a_{-n}\a_n -\at_{-n}\at_n )
+\right.\nonumber\\
&+& \left. 
\frac{1}{N}\( \sum_{m,n>0}+\sum_{m,n<0} \)
 :\a_m\a_n\a_{-m-n}+\at_m\at_n\at_{-m-n} : 
\] \, .
\ees
which is exactly \eq{HUN} as we expected. Finally it is obvious that,
being $H_{U(N)}$ and $Q$ the same ones of the $\l = 1$ case,
$H_{SU(N)}$ is given by \eq{HSUN}.

\end{document}